\documentclass[onecolumn,secnumarabic,amssymb, nobibnotes, aps, prd,showpacs]{revtex4}
\usepackage[utf8]{inputenc}
\usepackage{amssymb,amsmath,graphicx,latexsym}
\usepackage{array}
\usepackage{bm}
\usepackage{color}
\usepackage{dcolumn}
\usepackage{multirow}
\usepackage{float}

\newcommand{\dtabsize}{\scriptsize}
\newcommand{\lrt}{\leftrightarrow}
\begin{document}
%\tiny         size 5pt
%\scriptsize   size 7pt
%\footnotesize size 8pt
%\small        size 9pt
%\normalsize   size 10pt
%\large        size 12pt
%\Large        size 14.4pt
%\LARGE        size 17.28pt
%\huge         size 20.74pt
%\Huge         size 24.88pt

\title{Lorentz invariance violation in the neutrino sector: a joint analysis from big bang nucleosynthesis and the cosmic microwave background}

\author{Wei-Ming Dai$^{1,3}$}
\email{daiwming@itp.ac.cn}
\author{Zong-Kuan Guo$^{1,2}$}
\email{guozk@itp.ac.cn}
\author{Rong-Gen Cai$^{1,3}$}
\email{cairg@itp.ac.cn}
\author{Yuan-Zhong Zhang$^{1}$}
\email{zyz@itp.ac.cn}

\affiliation{
$^{1}$CAS Key Laboratory of Theoretical Physics, Institute of Theoretical Physics,
Chinese Academy of Sciences, P.O. Box 2735, Beijing 100190, China \\
$^{2}$School of Astronomy and Space Science, University of Chinese Academy of Sciences,
No.19A Yuquan Road, Beijing 100049, China \\
$^{3}$School of Physical Sciences, University of Chinese Academy of Sciences,
No.19A Yuquan Road, Beijing 100049, China}

\begin{abstract}
We investigate constraints on Lorentz invariance violation in the neutrino sector
from a joint analysis of big bang nucleosynthesis and the cosmic microwave background.
The effect of Lorentz invariance violation during the epoch of big bang nucleosynthesis
changes the predicted helium-4 abundance, which influences the power spectrum of the cosmic microwave background at the recombination epoch.
In combination with the latest measurement of the primordial helium-4 abundance,
the Planck 2015 data of the cosmic microwave background anisotropies
give a strong constraint on the deformation parameter
since adding the primordial helium measurement breaks the degeneracy
between the deformation parameter and the physical dark matter density.
\end{abstract}
\maketitle

\section{Introduction}
\label{s:introduction}

Neutrino oscillations have shown that there are small but nonzero mass squared differences
between three neutrino mass eigenstates (see Ref.~\cite{GonzalezGarcia:2007ib} and the reference therein),
which imply the existence of physics beyond the standard model of particle physics.
However, neutrino oscillation experiments cannot tell us the overall mass scale of neutrinos.
Fortunately, cosmology provides a promising way to determine or constrain the total mass of neutrinos
by the gravitational effects of massive neutrinos which can significantly change
the CMB power spectrum due to lensing~\cite{Kaplinghat:2003bh, Lesgourgues:2005yv, Lesgourgues:2012uu, Ade:2015xua}
and the formation of large-scale structures~\cite{Hu:1997mj, Ringwald:2004np},
therefore, alter the cosmic microwave background (CMB) anisotropies
and the large-scale structure distribution of matter
(see~\cite{Wong:2011ip} for a review).
Recent Planck 2015 data combined with some low-redshift data give an upper limit on the total mass
of neutrinos at $95\%$ confidence level, $\Sigma m_\nu<0.19$ eV~\cite{Ade:2015xua}.

Another possible signal of new physics is Lorentz invariance violation in the neutrino sector.
The observed neutrino oscillations may originate from a combination of effects involving neutrino
masses and Lorentz invariance violation~\cite{Coleman:1997xq,Coleman:1998ti,Kostelecky:2003xn,Kostelecky:2003cr,Kostelecky:2011gq,Motie:2012qj}.
Lorentz invariance is a fundamental symmetry in the standard model of particle physics.
Although present experiments have confirmed Lorentz invariance to a good precision~\cite{Diaz:2013wia},
it may be broken in the early Universe when energies approach the Planck scale.
The standard model itself is believed to be a low-energy effective theory of an underlying unified theory.
Lorentz invariance violation has been explored in quantum gravity~\cite{AmelinoCamelia:2002dx},
loop quantum gravity~\cite{Alfaro:1999wd}, non-commutative field theory~\cite{Carroll:2001ws},
and doubly special relativity theory~\cite{Magueijo:2001cr}.
Various searches for Lorentz invariance violation have been performed
with a wide range of systems~\cite{Colladay:1998fq, Kostelecky:2008ts}.

The cosmological consequences of neutrinos with the Coleman-Glashow type~\cite{Coleman:1998ti} dispersion relation
have been studied in Ref.~\cite{Guo:2012mv}.
As can be expected, the Lorentz-violating term affects not only the evolution of the cosmological background
but also the behavior of the neutrino perturbations.
The former changes the expansion rate prior to and during the epoch of photon-baryon decoupling,
which alters heights of the first and second peaks of the CMB temperature power spectrum,
while the latter alters the shape of the CMB power spectrum by changing neutrino propagation.
Since these two effects can be distinguished from a change in the total mass of neutrinos
or in the effective number of extra relativistic species,
CMB data have been proposed as a probe of Lorentz invariance violation in the neutrino sector.

Moreover, since the Lorentz-violating term influences the abundances of the light elements
by altering the energy density of the Universe and weak reaction rates
prior to and during the Big bang nucleosynthesis (BBN) epoch,
BBN provides a promising probe of Lorentz invariance violation
in the neutrino sector in the early Universe~\cite{Guo:2013zwa}.
In particular the BBN-predicted abundance of helium-4 is very sensitive to the Lorentz-violating term.
It is well known that the primordial helium-4 abundance plays a non-negligible role at the recombination epoch
because it has an influence on the number density of free electron.
Therefore, Lorentz invariance violation affects the CMB power spectrum
by changing the BBN-predicted abundance of helium-4.
A reasonable way to test Lorentz invariance violation in the neutrino sector with CMB data
is to take account of the BBN-predicted helium-4 abundance as a prior
rather than fixing the helium-4 abundance.

In this paper we investigate constraints on Lorentz invariance violation in the neutrino sector
from a joint analysis of BBN and CMB.
In combination with the latest measurement of the primordial helium-4 abundance,
the Planck 2015 data of the CMB anisotropies give a strong constraint on the deformation parameter.
Adding the primordial helium measurement can break effectively the degeneracy
between the deformation parameter and the physical dark matter density.

This paper is organized as follows. In Sect.~\ref{s:deformation} we parameterize Lorentz invariance violation in the neutrino sector.
In Sect.~\ref{s:BBN}, we derive the weak reaction rates in the Lorentz-violating extension of the standard model
and calculate the BBN prediction of the helium-4 abundance.
In Sect.~\ref{s:perturbation}, we derive the Boltzmann equation for neutrinos in the synchronous gauge
and calculate the CMB power spectrum.
In Sect.~\ref{s:constraints}, we place constraints on the deformation parameter
using the Planck 2015 data in combination with the latest measurement of the primordial helium-4 abundance.
Sect.~\ref{s:conclusion} is devoted to conclusions.

\section{Deformed dispersion relation}
\label{s:deformation}

At a phenomenological level, the deformed dispersion relation can be characterized by a power series of momentum.
In this paper, we consider a simple instance of the dispersion relation for neutrinos constructed in~\cite{Coleman:1998ti},
which can be parameterized as follows:
\begin{equation}
E^2=m^2+p^2+\xi p^2,
\label{eq:mass-energy}
\end{equation}
where $E$ is the neutrino energy, $m$ the neutrino mass, $p=\left(p^ip_i\right)^{1/2}$ the magnitude of the 3-momentum,
and $\xi$ the deformation parameter characterizing the Lorentz symmetry violation.

In the spatially flat Friedmann-Lemaintre-Robertson-Walker metric with the scale factor $a$,
the number density $n_\nu$, energy density $\rho_\nu$ and pressure $P_\nu$ for neutrinos with~\eqref{eq:mass-energy} are given by
\begin{eqnarray}
n_\nu&=&\frac{1}{a^3}\int \frac{d^3q}{\left(2\pi\right)^3}f_0\!\left(q\right)\,,
\label{eq:number}\\
\rho_\nu&=&\frac{1}{a^4}\int \frac{d^3q}{\left(2\pi\right)^3}\epsilon f_0\!\left(q\right)\,,
\label{eq:energy}\\
P_\nu&=&\frac{1}{3a^4}\int \frac{d^3q}{\left(2\pi\right)^3}\frac{\left(1+\xi\right)q^2}{\epsilon}f_0\!\left(q\right)\,.
\label{eq:pressure}
\end{eqnarray}
The phase space distribution for neutrinos is $f_0(q)=g_s\left[1+\exp\left(\epsilon/T_0\right)\right]^{-1}$,
where $g_s=2$ is the number of spin degrees of freedom,
$T_0$ the neutrino temperature today,
$q$ the magnitude of the comoving 3-momentum, and $\epsilon=\sqrt{m^2a^2+\left(1+\xi\right)q^2}$ the comoving energy.

%------------------------------------------------------------------
\section{BBN prediction}
\label{s:BBN}

The abundances of the light elements produced during the BBN epoch
depend on the competition between the nuclear and weak reaction rates and the expansion rate of the Universe.
In the standard cosmological scenario and in the framework of the electroweak standard model,
the dynamics of this phase is controlled by only one free parameter, the baryon
to photon number density.

We consider $N$ species of nuclides whose abundances $X_i$ are the number densities $n_i$
normalized with respect to the baryon number density $n_B$,
\begin{equation}
X_i=\frac{n_i}{n_B}\,,~~~~~~i=n,p,{}^2H,\cdots.
\label{eq:abundances}
\end{equation}
Their evolutions are ruled by the following Boltzmann equations,
\begin{equation}
\dot{X_i}=\sum_{j,k,l}N_i\left(\Gamma_{kl\rightarrow ij}\frac{X^{N_l}_l\, X^{N_k}_k}{N_l!\, N_k!}-\Gamma_{ij\rightarrow kl}\frac{X^{N_i}_i\, X^{N_j}_j}{N_i!\, N_j!}\right)\,,
\end{equation}
where $\Gamma$ denotes the reaction rate
and $N_i$ is the number of nuclide $i$ involved in the reaction.

In the case of Lorentz invariance violation in the neutrino sector,
the changes of the reaction rates of the following weak reactions have to be taken into account:
\begin{eqnarray}
\label{eq:nnpe}
n+\nu_e &\lrt& p+e^-\,,\\
n+e^+ &\lrt& p+\bar{\nu}_e\,,\\
n &\lrt& p+\bar{\nu}_e+e^-\,,
\label{eq:npee}
\end{eqnarray}
which determine the neutron-to-proton ratio when the baryons become uncoupled from the leptons.
The abundances of the light elements depend on the neutron-to-proton ratio at the onset of BBN.
As an example, let us compute the reaction rate of the process~\eqref{eq:nnpe}
in the Lorentz-violating extension of the standard model~\cite{Colladay:1998fq,Lambiase:2005kb}.
The reaction rate is~\cite{Guo:2013zwa}
\begin{equation}
\label{eq:gamma}
\Gamma=\left[1-\frac{3}{8}\xi-\frac{3\left(C^2_V-C^2_A\right)}{4\left(C^2_V+3C^2_A\right)}\xi\right]\left(1+\xi\right)^{-3/2}\Gamma^{\left(0\right)}\,,
\end{equation}
where $C_V$ and $C_A$ are the vector and axial coupling of the nucleon,
$\Gamma^{\left(0\right)}$ is the standard reaction rate per incident nucleon derived in Ref.~\cite{Lopez:1998vk}.
The first prefactor on the right-hand side of Eq.~\eqref{eq:gamma}
arises from the neutrino propagator and the $e\nu W$ coupling in the Lorentz-violating extension of standard model.
The second prefactor comes from the statistical distribution for neutrinos.
As for the other processes in \eqref{eq:nnpe}-\eqref{eq:npee},
their reaction rates can be simply derived by properly changing the statistical factors
and the delta function determined by the energy conservation for each reaction.
Therefore, the corrections to the conversion rate of neutron into proton
and its inverse rate are the same as in~\eqref{eq:gamma}.

In order to calculate the abundance of the light elements produced during the BBN epoch,
we modified the publicly available PArthENoPE code~\cite{Pisanti:2007hk}
to appropriately incorporate the Lorentz-violating term in the neutrino sector.
As shown in~\cite{Guo:2013zwa}, the BBN-predicted abundance of helium-4 is sensitive to the deformed parameter $\xi$
due to the fact that it is mainly determined by the neutron-to-proton ratio,
which is related to the weak reaction rate, neutrino number density and expansion rate of the Universe.
In our analysis we shall focus on the helium-4 abundance that influences the CMB power spectrum.

\begin{figure}
\begin{center}
\includegraphics[width=3.5in,height=2.5in]{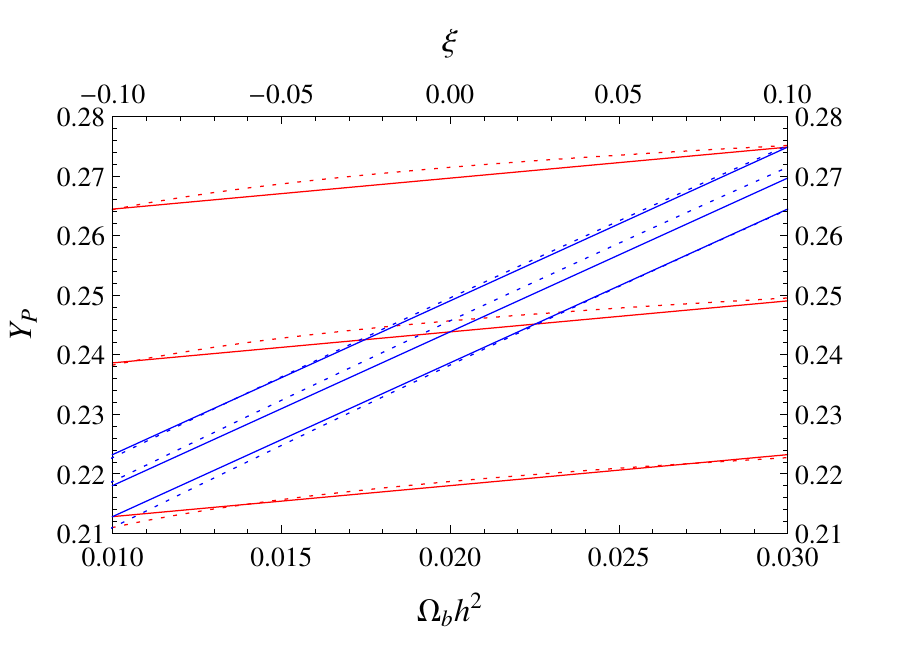}
\caption{BBN-predicted curves for $Y_\mathrm{p}$ in $\Omega_bh^2$ (red) and $\xi$ (blue).
From bottom to top, the red curves correspond to $\xi=-0.1$, $0$, $-0.1$,
while the blue curves correspond to $\Omega_bh^2=0.01$, $0.02$, $0.03$.
The dotted curves are the BBN-predicted results, while the solid curves
are our fits.}
\label{fig:Yp-xi-Omegabh2}
\end{center}
\end{figure}

Although in our analysis we use an interpolation method
to obtain the value of $Y_\mathrm{p}$ on a grid of points in $(\Omega_bh^2,\xi)$ space
(see Sect.~\ref{s:constraints}),
it is possible to describe the dependencies of the BBN-predicted helium-4 abundances upon the
two parameters by simple, linear fits which, over their ranges of applicability,
are accurate to a few percent.
Following~\cite{Kneller:2004jz}, we get the following linear fit for $Y_\mathrm{p}$ versus
$\Omega_bh^2$ and $\xi$:
\begin{equation}
Y_\mathrm{p}^\mathrm{FIT}=0.2334+0.52\Omega_bh^2+0.258\xi,
\end{equation}
over our adopted range in $\Omega_bh^2$ and $\xi$, as shown in Fig.~\ref{fig:Yp-xi-Omegabh2}.
We note that this fit is a very good approximation over the adopted parameter ranges,
which is useful in facilitating studies of the viability of various
options for non-standard physics and cosmology, prior to undertaking detailed BBN calculations

Since the neutrino species share the temperature with the photons prior to neutrino decoupling,
the prediction for the effective number of neutrino species is
\begin{equation}
N_\mathrm{eff}=3.046(1+\xi)^{-3/2}.
\end{equation}
The total relativistic energy density $\rho_r$ in neutrinos is given in terms of the
photon density $\rho_\gamma$ by
\begin{equation}
\rho_r=N_\mathrm{eff}\frac78\left(\frac{4}{11}\right)^{4/3}\rho_\gamma.
\end{equation}

%----------------------------------------------------------------------
\section{CMB power spectrum}
\label{s:perturbation}

The Lorentz-violating term affects not only the evolution of the cosmological background
but also the behavior of the neutrino perturbations.
Following Ref.~\cite{Ma:1995ey} we shall derive the Boltzmann equation of the distribution function
for neutrinos with~\eqref{eq:mass-energy} to linear order in the perturbations.
In the synchronous gauge,
the perturbed energy density $\delta\rho_\nu$, pressure $\delta P_\nu$,
energy flux $\theta_\nu$, and shear stress $\sigma_\nu$ are, respectively, given by
\begin{eqnarray}
\delta\rho_\nu&=&\frac{1}{a^4}\int \frac{d^3q}{(2\pi)^3} \epsilon f_0\!\left(q\right)\Psi_0\,,
\label{eq:denergy}\\
\delta P_\nu&=&\frac{1}{3a^4}\int \frac{d^3q}{(2\pi)^3} \frac{\left(1+\xi\right)q^2}{\epsilon} f_0\!\left(q\right)\Psi_0\,,
\label{eq:dpresssure}\\
(\rho_\nu+P_\nu)\theta_\nu &=&\frac{k}{a^4}\int \frac{d^3q}{(2\pi)^3} \sqrt{1+\xi}q f_0\!\left(q\right)\Psi_1\,,
\label{eq:dflux}\\
(\rho_\nu+P_\nu) \sigma_\nu &=&\frac{2}{3a^4}\int \frac{d^3q}{(2\pi)^3} \frac{\left(1+\xi\right)q^2}{\epsilon} f_0\!\left(q\right)\Psi_2\,,
\label{eq:shear}
\end{eqnarray}
where $\Psi_l$ are the expansion coefficients of the perturbed neutrino distribution function
expanded in a series of Legendre polynomials
\begin{equation}
\Psi(\vec{k},\hat{n},q,\tau)=\sum_{\ell=0}^{\infty}\left(-\!i\right)^\ell\left(2\ell+1\right)\Psi_\ell(\vec{k},q,\tau)P_\ell (\hat{k}\cdot\hat{n})\,,
\label{eq:Legendre-mass}
\end{equation}
where $\tau$ is the conformal time. These perturbations $\Psi_l$ satisfy the following Boltzmann equations:
\begin{eqnarray}
&&\dot{\Psi}_\ell+\left(1+\xi\right)\frac{q}{\epsilon}\frac{k}{2\ell+1}\left[\left(\ell+1\right)\Psi_{\ell+1}-\ell \Psi_{\ell-1}\right] \nonumber \\
&& \quad +\left(\delta_{2\ell}\frac{1}{15}\dot{h}+\delta_{2\ell}\frac{2}{5}\dot{\eta}-\delta_{0\ell}\frac{1}{6}\dot{h}\right)\frac{d \ln f_0}{d \ln q}=0\,,
\label{eq:boltz-eqs-mass}
\end{eqnarray}
in the synchronous gauge, where dots denote derivatives with respect to the conformal time,
$h$ and $\eta$ are the two scalar modes of the metric perturbations in the Fourier space $k$.
To avoid the reflections from high-$\ell$ equations by simply set $\Psi_\ell=0$ for $\ell>\ell_{\rm max}$,
such a Boltzmann hierarchy is effectively truncated by adopting the following scheme~\cite{Ma:1995ey}:
\begin{equation}
\Psi_{\ell_{\rm max}+1}\approx\frac{\left(2\ell_{\rm max}+1\right)\epsilon}{\left(1+\xi\right)qk\tau}\Psi_{\ell_{\rm max}}-\Psi_{\ell_{\rm max}-1}\,.
\label{eq:boltz-trun-mass}
\end{equation}

Making a transformation $q\rightarrow\sqrt{1+\xi}q$,
we note that the number density, energy density and pressure of neutrinos are proportional to $(1+\xi)^{-3/2}$,
which means that increasing $\xi$ decreases the number density, energy density and pressure of neutrinos.
Moreover, there is a factor $\sqrt{1+\xi}$ in the second term of the Boltzmann equations,
which comes from neutrino propagation.

For the massless neutrinos, the Boltzmann equations are simplified by setting $\epsilon=\sqrt{1+\xi}q$.
The number of variables can be reduced by integrating out the $q$ dependence
in the neutrino distribution function so that
\begin{align}
\nonumber
F(\vec{k},\hat{n},\tau)&=\frac{\int q^2 dq \epsilon f_0(q) \Psi}{\int q^2 dq \epsilon f_0(q)}\\
&=\sum_{\ell=0}^{\infty}(-\!i)^\ell(2\ell+1)F_\ell (\vec{k},\tau)P_\ell (\hat{k} \cdot \hat{n})\,.
\label{eq:Legendre-massless}
\end{align}

Using the orthonormality and recursion relation of the Legendre polynomials,
from the Boltzmann equations we derive the evolution equations of
the perturbed energy density $\delta_\nu=\delta \rho_\nu/\rho_\nu$,
energy flux $\theta_\nu$, shear stress $\sigma_\nu$ independent of the momentum $q$,
\begin{eqnarray}
\dot{\delta}_\nu&=&-\!\frac{4}{3}\sqrt{1+\xi}\theta_\nu-\frac{2}{3}\dot{h}\,,
\label{eq:denergy-kmassless}\\
\dot{\theta}_\nu&=&\sqrt{1+\xi}k^2\left(\frac{1}{4}\delta_\nu-\sigma_\nu\right)\,,
\label{eq:dflux-kmassless}\\
\nonumber
2\dot{\sigma}_\nu &=& \sqrt{1+\xi}\frac{8}{15}\theta_\nu\\
&&-\frac{3}{5}\sqrt{1+\xi}k F_3+\frac{4}{15}\dot{h}+\frac{8}{5}\dot{\eta}\,,
\label{eq:shear-kmassless}\\
\dot{F}_\ell&=&\frac{\sqrt{1+\xi}k}{2\ell+1}\left[\ell F_{\ell-1}-\left(\ell+1\right)F_{\ell+1}\right]\,,  \ell\geq3\,.
\label{eq:recursion-kmassless}
\end{eqnarray}
The truncation scheme for massless neutrinos is
\begin{equation}
F_{\ell_{\rm max}+1}\approx\frac{\left(2\ell_{\rm max}+1\right)}{\sqrt{1+\xi}k\tau}F_{\ell_{\rm max}}-F_{\ell_{\rm max}-1}\,.
\label{eq:boltz-trun-massless}
\end{equation}

In the synchronous gauge the adiabatic initial conditions for the metric perturbations
and massless neutrinos in the cold dark matter frame are
\begin{eqnarray}
h &=& Ck^2\tau^2-\frac{1}{5}C\omega k^2 \tau^3\,,\\
\eta &=& 2C-\frac{5-5R_\nu+9 R_\nu\sqrt{1+\xi}}{6\left(15+4R_\nu\sqrt{1+\xi}\right)}C k^2\tau^2\,,\\
\delta_\nu &=& -\!\frac{2}{3}C k^2\tau^2+\frac{2}{15}C\omega k^2\tau^3\,,\\
\theta_\nu &=& -\!\frac{\left(23+4R_\nu\right)\sqrt{1+\xi}}{18\left(15+4R_\nu\sqrt{1+\xi}\right)}C k^4\tau^3\,,\\
\nonumber
\sigma_\nu &=& \frac{2\left(2+R_\nu-R_\nu\sqrt{1+\xi}\right)}{3\left(15+4R_\nu\sqrt{1+\xi}\right)}C k^2\tau^2 \\
\nonumber
&-& \frac{25+R_\nu\left(65-85\sqrt{1+\xi}\right)+4R_\nu^2\left(\sqrt{1+\xi}-1-\xi\right)}
  {15\left(15+2R_\nu\sqrt{1+\xi}\right)\left(15+4R_\nu\sqrt{1+\xi}\right)} \\
  &\times& C\omega k^2\tau^3\,,\\
F_3 &=& \frac{4}{21}\frac{2\sqrt{1+\xi}+R_\nu\left(\sqrt{1+\xi}-1-\xi\right)}{15+4R_\nu\sqrt{1+\xi}}C k^3\tau^3\,,
\end{eqnarray}
where $C$ is a dimensionless constant determined by the amplitude of the fluctuations from inflation,
$R_\nu \equiv \rho_\nu/(\rho_\gamma+\rho_\nu)$,
and $\omega \equiv \Omega_m H_0/\sqrt{\Omega_\gamma+\Omega_\nu}$,
which corresponds to the matter contribution to the total energy density of the Universe.
In the radiation-dominated era the massive neutrinos are relativistic.
The initial condition of $\Psi_0$ for massive neutrinos is related to $\delta_\nu$.
Then, using Eqs.~\eqref{eq:boltz-eqs-mass} and
ignoring the mass terms in the differential equations of high-order moment,
we get the initial conditions of $\Psi_\ell$ for massive neutrinos:
\begin{eqnarray}
\Psi_0&=&-\!\frac{1}{4}\delta_\nu\frac{d \ln f_0}{d \ln q}\,,\\
\Psi_1&=&-\!\frac{\epsilon}{3\sqrt{1+\xi}q k}\theta_\nu \frac{d \ln f_0}{d \ln q}\,,\\
\Psi_2&=&-\!\frac{1}{2(1+\xi)}\left(\sigma_\nu+\frac14 \xi \delta_\nu\right)\frac{d \ln f_0}{d \ln q}\,,\\
\Psi_\ell&=&0,~~ \ell\ge3\,.
\end{eqnarray}

\begin{figure}
\begin{center}
\includegraphics[width=3.5in,height=2.5in]{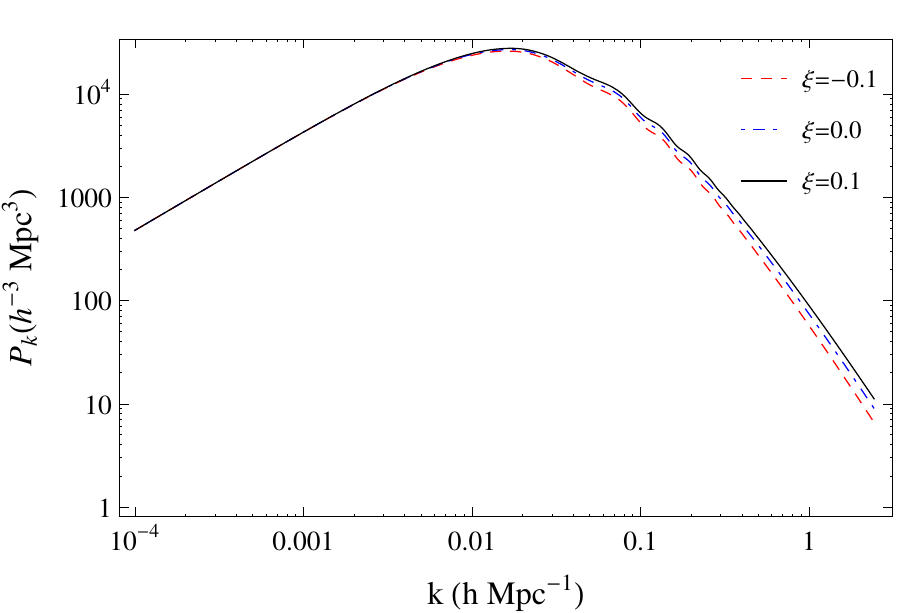}
\caption{The matter power spectrum for $\xi=-0.1$ (dashed), $0$ (dash-dotted), $0.1$ (solid).}
\label{fig:mspectra-xi}
\end{center}
\end{figure}

In order to compute the theoretical CMB power spectrum,
we modified the Boltzmann CAMB code~\cite{Lewis:1999bs} to appropriately incorporate the Lorentz-violating term in the neutrino sector.
As pointed out in~\cite{Guo:2012mv}, the effects of the Lorentz-violating term on
the CMB power spectrum are distinguished from a change in the
total mass of neutrinos or in the effective number of extra relativistic
species~\cite{Hu:1995fqa,Bashinsky:2003tk,Hamann:2007pi,Hou:2011ec,Guo:2012jn}.
Therefore, the measurements of the CMB anisotropies provide a cosmological probe of
Lorentz invariance violation in the neutrino sector.
Moreover, we calculate the matter power spectrum today for $\xi=-0.1, 0, 0.1$.
From Fig.~\ref{fig:mspectra-xi} we find that increasing $\xi$ can enhance the matter power spectrum at small scales.
It is well known that variances in the helium-4 abundance modify the density of free electrons
between helium and hydrogen recombination and therefore influence the CMB power spectrum.
However, the effects of the Lorentz-violating term on the BBN-predicted helium-4 abundance
were not considered in~\cite{Guo:2012mv}.
In the next section, we shall put constraints on the deformation parameter
from a joint analysis of BBN and CMB.

%----------------------------------------------------------------------
\section{Joint analysis from BBN and CMB}
\label{s:constraints}

In our analysis we use a modified version of the publicly available CosmoMC package
to explore the parameter space by means of Monte Carlo Markov chains technique~\cite{Lewis:2002ah}.
Since the deformation parameter $\xi$ is nearly uncorrelated with the total mass of neutrinos
when CMB data are used to detect the signature of Lorentz invariance violence~\cite{Guo:2012mv},
we only consider massless neutrinos with vanishing chemical potentials in our analysis.
Our cosmological model is the spatially flat $\Lambda$CDM plus
three types of massless neutrino with the deformed dispersion relation~\eqref{eq:mass-energy},
which can be described by the following seven parameters:
\begin{equation*}
\{\Omega_bh^2,\Omega_ch^2,\theta_\mathrm{MC},\tau_{\rm re},n_s,A_s,\xi\}\,,
\end{equation*}
where $h$ is the dimensionless Hubble parameter defined by $H_0=100h$~km~$\rm s^{-1}$~$\rm Mpc^{-1}$,
$\Omega_bh^2$ and $\Omega_ch^2$ are the physical baryon and dark matter densities relative to the critical density,
$\theta_\mathrm{MC}$ is an approximation to the ratio of the sound horizon to the angular diameter distance at the photon decoupling,
$\tau_{\rm re}$ is the reionization optical depth,
$n_s$ and $A_s$ are the spectral index and amplitude of the primordial curvature perturbations at the pivot scale $k_0=0.05$ Mpc$^{-1}$.

We firstly run the PArthENoPE BBN code with two free input parameters, $\Omega_bh^2$ and $\xi$,
to precalculate the primordial helium-4 abundance on a grid of points in $(\Omega_bh^2,\xi)$ space.
For each couple of parameter values the helium-4 abundances are then obtained from the grid by
two-dimensional cubic spline interpolation,
which is adopted to calculate the CMB power spectrum.
This method is called ``BBN consistency".

\begin{table*}
\begin{center}
\caption{Mean values and marginalized 68\% confidence level for the deformation parameter and other cosmological parameters,
derived from the Planck data in combination with baryon acoustic oscillation data, the JLA sample of Type Ia supernovae
and the measurement of the helium-4 mass fraction. For comparison with BBN consistency, fixing $Y_{\rm p}=0.24$ is also considered. }
\begin{tabular}{|>{\dtabsize}c|>{\dtabsize}c|>{\dtabsize}c|>{\dtabsize}c|}
\hline
\hline
\multirow{2}{*}{Model}      &\multicolumn{2}{>{\dtabsize}c|}{Planck2015+JLA+BAO} &Planck2015+JLA+BAO+$Y_{\rm p}$ \\ \cline{2-4}
                                &$Y_{\rm p}=0.24$ (fixed)           &BBN consistency         &BBN consistency         \\
\hline
$\Omega_{\mathrm{b}}h^2$        &$0.02127\pm0.00046$     &$0.02212\pm0.00025$     &$0.02227\pm0.00021$   \\
$\Omega_{\mathrm{c}} h^2$       &$0.1124\pm0.0061$       &$0.1163\pm0.0040$       &$0.1197\pm0.0018$     \\
$100\theta_{\mathrm{MC}}$       &$1.04181\pm0.00098$     &$1.04173\pm0.00101$     &$1.04090\pm0.00050$   \\
$\tau_{\rm re}$                 &$0.071\pm0.018$         &$0.076\pm0.018$         &$0.077\pm0.018$       \\
$n_\mathrm{s}$                  &$0.9583\pm0.0069$       &$0.9679\pm0.0049$       &$0.9659\pm0.0044$     \\
$\ln(10^{10} A_\mathrm{s})$     &$3.064\pm0.038$         &$3.084\pm0.035$         &$3.087\pm0.035$       \\
$H_{\mathrm 0}$(km $\mathrm s^{-1}$$\rm Mpc^{-1}$)        &$65.92\pm1.76$          &$67.25\pm1.00$          &$68.01\pm0.63$        \\
$\sigma_\mathrm{8}$             &$0.840\pm0.014$         &$0.846\pm0.016$         &$0.841\pm0.015$       \\
\rm{Age(Gyr)}                   &$14.191\pm0.35$         &$13.937\pm0.19$         &$13.767\pm0.068$      \\
$\Omega_\mathrm{\Lambda}$       &$0.691\pm0.008$         &$0.694\pm0.007$         &$0.693\pm0.007$       \\
$\Omega_\mathrm{m}$             &$0.309\pm0.008$         &$0.306\pm0.007$         &$0.307\pm0.007$       \\
$\xi$                           &$0.101\pm0.091$         &$0.042\pm0.049$         &$-0.002\pm0.014$      \\
\hline
\end{tabular}
\label{table:parameters}
\end{center}
\end{table*}

We use the recently released Planck 2015 likelihood code and data, including the Planck low-$\ell$ likelihood
at multipoles $2\le \ell \le 29$ and Planck high-$\ell$ likelihood at multipoles $\ell \ge 30$
based on pseudo-$C_\ell$ estimators~\cite{Ade:2015xua}.
The former uses foreground-cleaned LFI $70$ GHz polarization maps together
with the temperature map obtained from the Planck $30$ to $353$ GHz channels
by the Commander component separation algorithm over 94\% of the sky.
The latter uses $100$, $143$, and $217$ GHz half-mission cross-power spectra,
avoiding the galactic plane as well as the brightest point sources and the regions
where the CO emission is the strongest.
“Planck 2015” denotes the combination of the low-$\ell$ temperature-polarization likelihood
and the high-$\ell$ temperature likelihood.

We use the “Joint Light-curve Analysis” (JLA) sample of type Ia supernovae,
which is constructed from the SNLS and SDSS supernova data,
together with several samples of low-redshift supernovae~\cite{Betoule:2012an}.
Baryon acoustic oscillation (BAO) measurements are another important astrophysical data set,
which are powerful to break parameter degeneracies from CMB measurements.
We use BAO measurements of $D_V/r_{\rm drag}$ from the 6dFGS at $z_{\rm eff}=0.106$,
SDSS Main Galaxy Sample at $z_{\rm eff}=0.15$, BOSS LOWZ at $z_{\rm eff}=0.32$,
and BOSS CMASS at $z_{\rm eff}=0.57$~\cite{Ade:2015xua}.
Here $D_V$ is the effective distance measure for angular diameter distance,
$r_{\rm drag}$ is the comoving sound horizon at the end of the baryon drag epoch
and $z_{\rm eff}$ is the effective redshift.
In our analysis we also use the helium-4 mass fraction measurement of $Y_{\rm p}=0.2449\pm0.0040$,
derived from helium and hydrogen emission lines from metal-poor extragalactic HII regions
and from a regression to zero metallicity~\cite{Aver:2015iza}.

\begin{figure}
\begin{center}
\includegraphics[width=3.5in,height=2.5in]{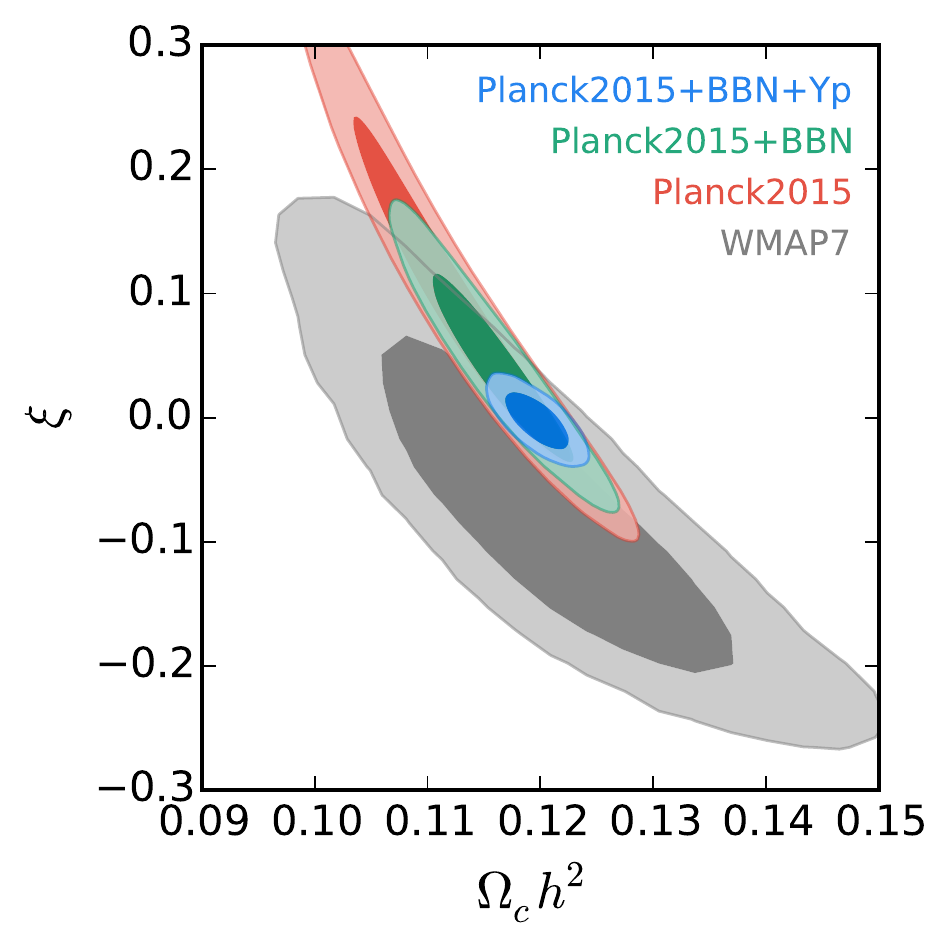}
\caption{Marginalized two-dimensional contours in the $\xi-\Omega_c h^2$ plane
with $68\%$ and $95\%$ confidence level in the extended $\Lambda$CDM model with BBN consistency,
derived from the data of Planck 2015+JLA+BAO (green) and
Planck 2015+JLA+BAO+$Y_{\rm p}$ (blue).
The red and gray contours correspond to $Y_{\rm p}=0.24$, derived from Planck 2015+JLA+BAO
and from WMAP7+$H_0$+BAO~\cite{Guo:2012mv}, respectively.}
\label{fig:xi-omegach2}
\end{center}
\end{figure}

\begin{figure}
\begin{center}
\includegraphics[width=3.2in,height=2.5in]{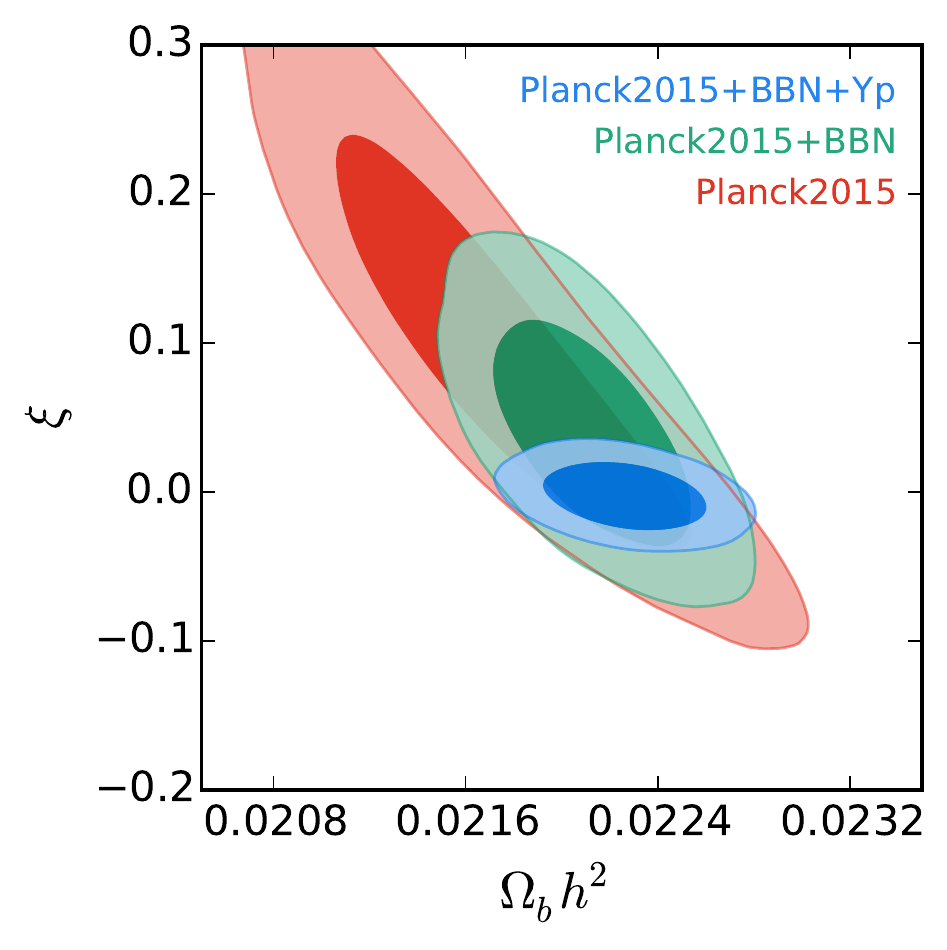}
\includegraphics[width=3.2in,height=2.5in]{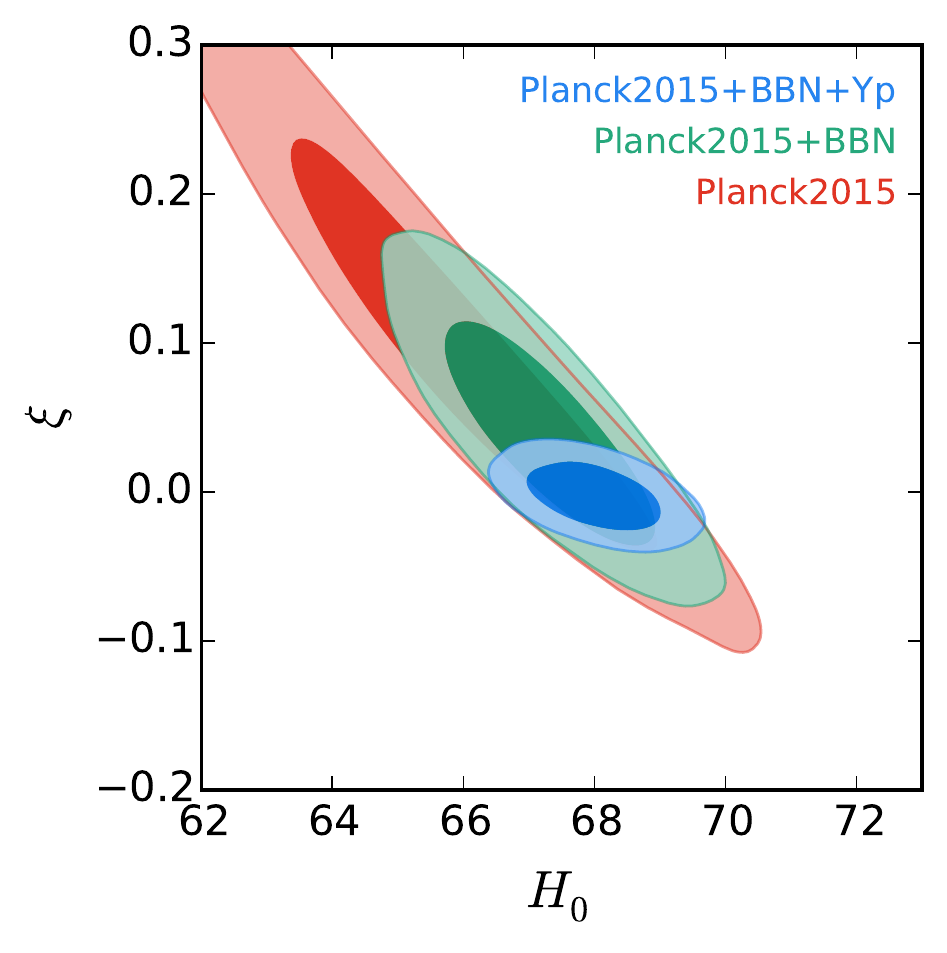}
\includegraphics[width=3.2in,height=2.5in]{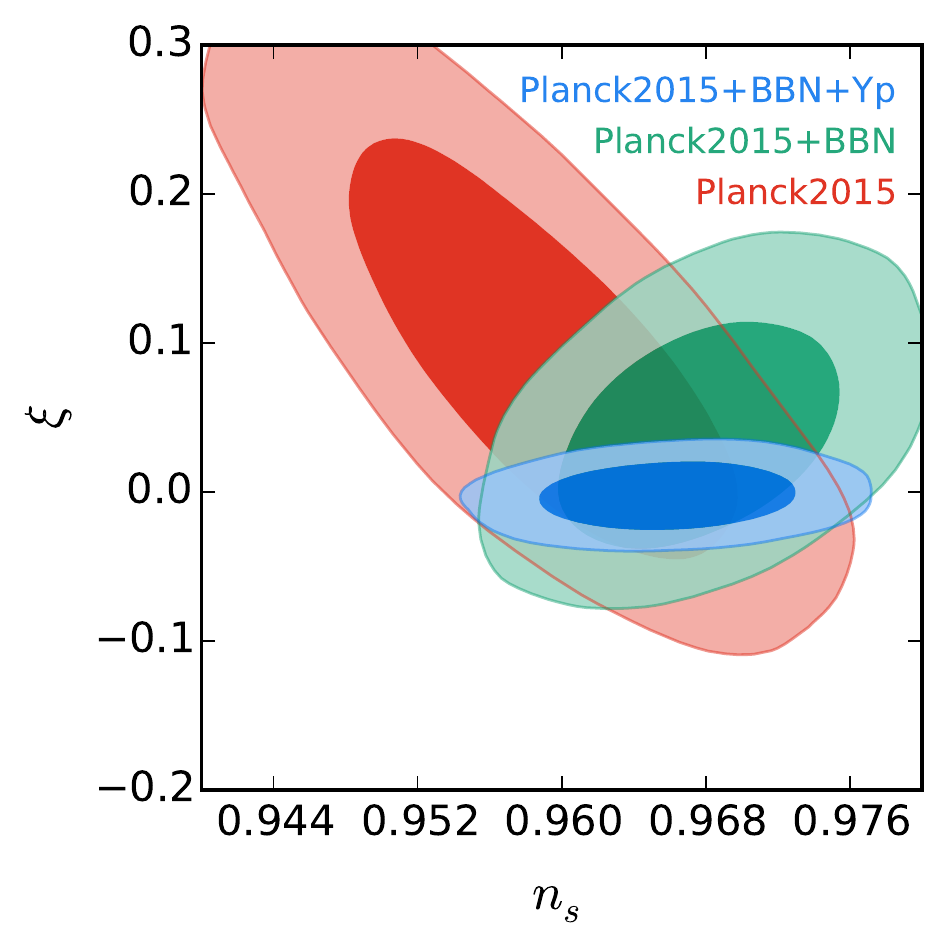}
\includegraphics[width=3.2in,height=2.5in]{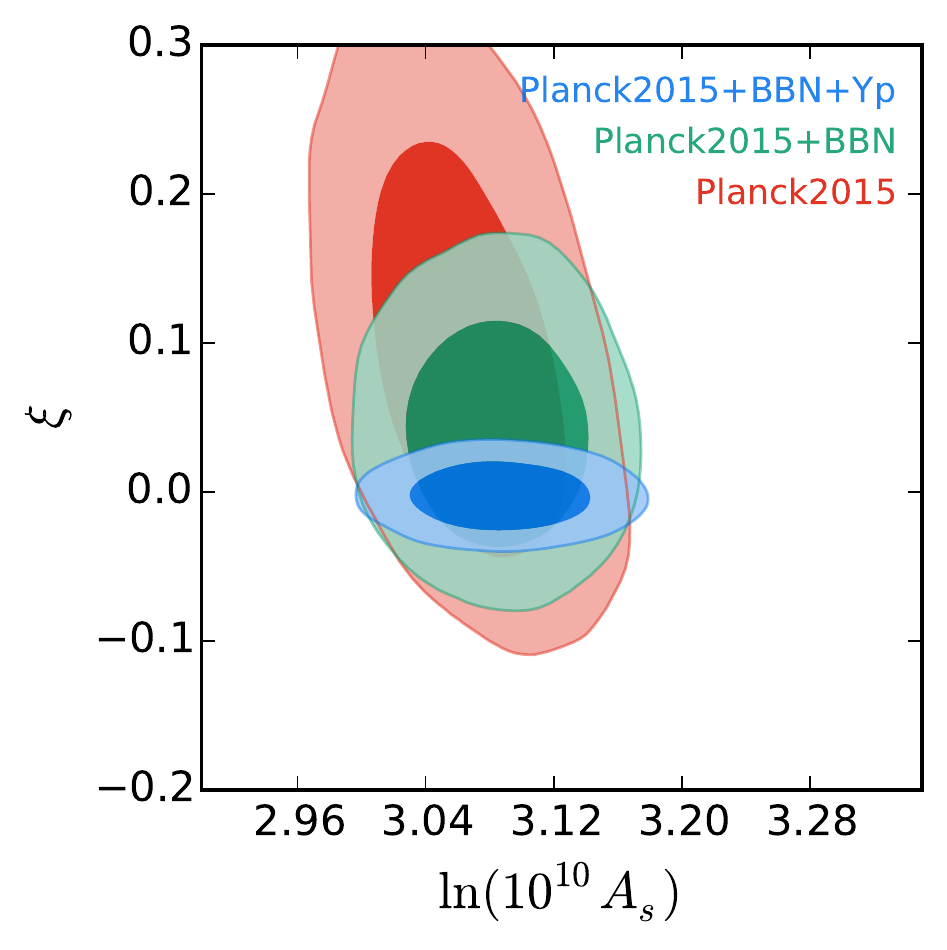}
\caption{Marginalized two-dimensional contours in $\xi-\Omega_b h^2$ (top-left),
$\xi-H_0$ (top-right), $\xi-n_s$ (bottom-left) and $\xi-A_s$ (bottom-right),
with $68\%$ and $95\%$ confidence level in the extended $\Lambda$CDM model with BBN consistency,
derived from the data of Planck 2015+JLA+BAO (green) and
Planck 2015+JLA+BAO+$Y_{\rm p}$ (blue).
The red contours correspond to $Y_{\rm p}=0.24$, derived from Planck 2015+JLA+BAO.}
\label{fig:xi-others}
\end{center}
\end{figure}

In Table~\ref{table:parameters} we list our results.
With the combined data of Planck 2015+JLA+BAO, the deformation parameter is estimated to be $\xi=0.042\pm 0.049$
if the helium-4 abundance is set from the BBN prediction.
For comparison, if the helium-4 abundance is not varied independently of other parameters,
with the same data set we find $\xi=0.101\pm 0.091$,
which indicates that the constraint of $\xi$ is not improved compared to the result
derived in~\cite{Guo:2012mv} from the 7-year WMAP~\cite{Komatsu:2010fb}
in combination with lower-redshift measurements of the expansion rate.
The reason is that the estimated values of cosmological parameters are biased by fixing $Y_{\rm p}=0.24$.
The gray and red regions in Fig.~\ref{fig:xi-omegach2} show 68\% and 95\% contours
in the $\xi–\Omega_c h^2$ plane setting $Y_{\rm p}=0.24$, derived from Planck 2015+JLA+BAO
and from WMAP7+$H_0$+BAO~\cite{Guo:2012mv}, respectively.
We see that the Planck 2015 data marginally favor a positive $\xi$
compared to the 7-year WMAP data.
The addition of the latest measurement of the helium-4 abundance leads to
a strong constraint on the deformation parameter $\xi=-0.002\pm 0.014$.
As shown in Figure~\ref{fig:xi-omegach2}, adding the $Y_{\rm p}$ data
breaks the degeneracy between the deformation parameter and the physical dark matter density.
This result shows that no signal of Lorentz invariant violation is detected
by the joint analysis of CMB and BBN.
Figure~\ref{fig:xi-others} shows the constraints in the $\xi-\Omega_bh^2$ plane,
$\xi-H_0$ plane, $\xi-n_s$ plane and $\xi-A_s$ plane from Planck 2015+JLA+BAO.

%----------------------------------------------------------------------
\section{Conclusions and discussions}
\label{s:conclusion}

We have studied the cosmological constraints on Lorentz invariance violation
in the neutrino sector by a joint analysis of CMB and BBN.
Instead of fixing the value of the helium-4 abundance,
we have applied the BBN prediction of the helium-4 abundance
determined by $\xi$ and $\Omega_b h^2$ to calculate the CMB power spectra.
Using the Planck 2015 data in combination with the JLA sample of type Ia supernovae
and the BAO feature, we have put a constraint on the deformation parameter.
Adding the measurement of the helium-4 abundance breaks the degeneracy
between the deformation parameter and the physical dark matter density
and so improves the constraint by a factor 3.

\begin{figure}
\begin{center}
\includegraphics[width=3.5in,height=2.5in]{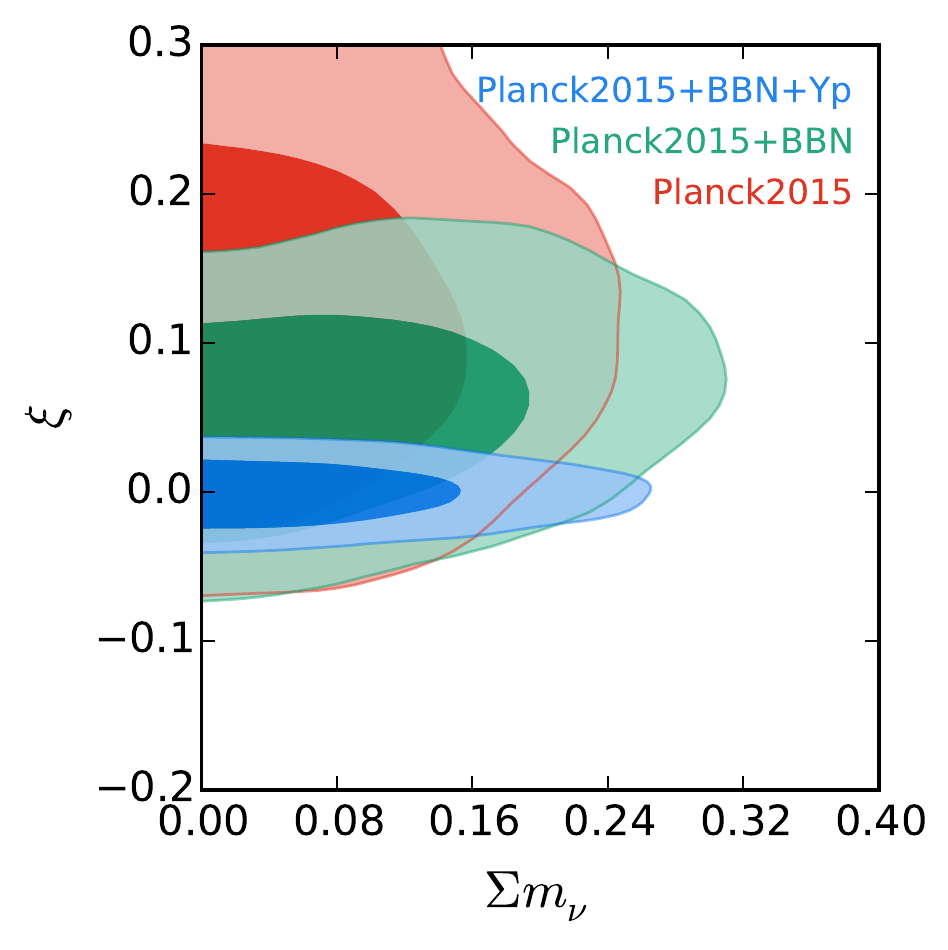}
\caption{Marginalized two-dimensional contours in the $\xi-\Sigma m_\nu$ plane
with $68\%$ and $95\%$ confidence level in the extended $\Lambda$CDM model with BBN consistency,
derived from the data of Planck 2015+JLA+BAO (green) and
Planck 2015+JLA+BAO+$Y_{\rm p}$ (blue).
The red contours correspond to $Y_{\rm p}=0.24$, derived from Planck 2015+JLA+BAO.}
\label{fig:xi-mnu}
\end{center}
\end{figure}

As found in~\cite{Guo:2012mv}, the deformation parameter is nearly uncorrelated with the total mass of neutrinos $\Sigma m_\nu$
when WMAP data are used to constrain $\xi$.
Planck 2015 data is much more sensitive to the total mass of neutrinos than WMAP data.
Therefore, we consider the effects of the neutrino masses on our results.
Figure~\ref{fig:xi-mnu} shows 68 and 95\% contours in the $\xi-\Sigma m_\nu$ plane
for Planck 2015 data combined with lower-redshift measurements.
We see that $\xi$ is nearly uncorrelated with $\Sigma m_\nu$ from Fig.~\ref{fig:xi-mnu}.

Compared to cosmological bounds on the Lorentz-violating coefficient,
observations of high-energy astrophysical neutrinos give stronger constraints.
As listed in Table XIII of Ref.~\cite{Kostelecky:2011gq},
the coefficient is constrained down to ${\cal O}(10^{-9})$ from the time-of-flight measurements,
under the assumption that neutrino oscillations are negligible.
Cohen and Glashow have argued that the observation of neutrinos
with energies in excess of 100 TeV and a baseline of at
least 500 km allows us to deduce that the Lorentz-violating
parameter is less than ${\cal O}(10^{-11})$~\cite{Cohen:2011hx}.
The present work provides a new way to probe the signal of Lorentz invariance violation
in the early Universe, which can in principle be used to constrain $\xi$
in the sterile neutrino sector~\cite{Hamann:2010bk, Hamann:2011ge}.
Although present cosmological data are too weak to yield competitive constraints,
future measurements of CMB and BBN offer prospects for placing stringent constraints.

%----------------------------------------------------------------------------
\begin{acknowledgments}
Our numerical analysis was performed on the ``Era'' of Supercomputing Center,
Computer Network Information Center of Chinese Academy of Sciences.
This work is supported in part by the National Natural Science Foundation of China Grants No.11575272, No.11690021, No.11690022, No.11335012, No.11375247, No.11435006 and No.91436107, in part by the Strategic Priority Research Program of the Chinese Academy of Sciences, Grant No.XDB23030100 and No.XDB21010100,
and by Key Research Program of Frontier Sciences, CAS.
\end{acknowledgments}
%------------------------------------------------------------------------

\end{document}